\documentclass[aps,prl,twocolumn,epsfig,amsmath,amssymb]{revtex4}
\usepackage[english]{babel} \usepackage{latexsym}
\usepackage{graphicx} \usepackage{subfigure} \usepackage{epsfig}
\usepackage{psfrag}\usepackage{color}

\begin{document}

\title{Mott-insulator phases of non-locally coupled 1D dipolar Bose gases}
\author{A. Arg\"uelles and L. Santos} 
\affiliation{
\mbox{Institut f\"ur Theoretische Physik , Leibniz Universit\"at
Hannover, Appelstr. 2, D-30167, Hannover, Germany}\\
}

\begin{abstract}  
%
%

We analyze the Mott-insulator phases of dipolar bosonic gases placed in 
neighboring but unconnected 1D traps. Whereas for short-range interactions 
the 1D systems are independent, the non-local dipole-dipole interaction 
induces a direct Mott-insulator to pair-superfluid transition which 
significantly modifies the boundaries of the lowest Mott-insulator phases. 
The lowest boundary of the lowest Mott regions becomes progressively 
constant as a function of the hopping rate, eventually inverting its slope, leading 
to a re-entrant configuration which is retained in 2D. We discuss the consequences 
of this effect on the spatial Mott-insulator plateaux in experiments with 
additional harmonic confinement, showing that anti-intuitively 
the plateaux may become wider for increasing hopping. 
Our results are also applicable to non-dipolar boson-boson mixtures.

\end{abstract}  
 \maketitle


Strongly-correlated atomic gases have recently attracted 
a rapidly-growing attention, mainly motivated by impressive 
developments on the manipulation of atoms in optical lattices. 
When loaded in these lattices, ultra cold atoms experience a 
periodic potential that resembles that of electrons in solids, opening 
fascinating links between the physics of cold atoms 
and solid-state physics. In particular, cold bosons restricted to 
the lowest lattice band can be described by the 
Bose-Hubbard model \cite{Jaksch98}, which presents two different phases at zero temperature 
\cite{Fisher89}, namely a superfluid (SF) phase, and a gapped incompressible insulator phase 
known as Mott-insulator (MI), characterized by a commensurate occupation per lattice site. 
The SF to MI transition in cold bosons in optical lattices 
was recently observed in a remarkable experiment \cite{Mott}, in which the gapped nature of the 
MI excitation spectrum was clearly demonstrated. The realization of the MI 
was indeed possible due to an additional harmonic potential overimposed 
to the lattice that guaranteed locally the (otherwise practically impossible) commesurability 
condition necessary for the MI. Such inhomogeneous potential leads to the formation 
of spatial MI and SF shells \cite{Jaksch98,Batrouni}, which 
have been observed very recently \cite{BlochDomains,KetterleDomains}.

Among the various research lines related with optical lattices, the physics of 
mixtures has attracted a considerable attention. Bose-Fermi mixtures 
may lead to a wealth of phases of fermion composites \cite{BFMixt}, and 
may allow for the generation and engineering of disorder 
\cite{Sanpera,Castin,Esslinger,Sengstock}. Bose-Bose mixtures have also attracted a
major interest \cite{Altman,Kuklov,Ziegler,Mathey,Mishra}. In particular, it has 
been shown \cite{Kuklov} that the interspecies interaction may lead to the formation of 
a pair superfluid (PSF), i.e. a superfluid of boson-boson (or hole-hole) composites, which 
occurs in addition to MI phases for both components, as well as uncorrelated superfluid phases 
in each one of the components (2SF). 

Dipolar gases also attract currently a major interest, motivated by recent experiments 
on atoms with large magnetic moment \cite{Grismaier2005}, polar molecules \cite{Molecules} 
and Rydberg atoms \cite{Rydberg}. In these gases the long-range and anisotropic dipole-dipole 
interactions (DDI) become significant or even dominant when compared to the short-range 
isotropic interactions \cite{Dipoles}. The DDI can play an important role in 
the physics of lattice bosons, leading to additional 
phases, as checker-board or supersolid, which may be easily controllable by manipulating the atomic confinement 
\cite{Goral}. In addition, contrary to short-range interacting gases for which disconnected sites 
(i.e. without hopping between them) are fully independent, 
the long-range character of the DDI induces a coupling even for 
unconnected sites. The latter leads to fundamentally new physics, as e.g. a 
condensate of filaments \cite{Wang06}, novel quantum phase transitions in 
bilayer systems of polar molecules \cite{Wang06b}, 
or inelastic interlayer scattering for dipolar solitons \cite{Nath06}.

In this Letter, we show that the non-local DDI induces a direct MI-to-PSF transition for neighboring 
unconnected 1D dipolar gases, which significantly modifies the boundaries of the lowest 1D MI phases.
The same effect is also expected under appropriate conditions for non-dipolar boson-boson mixtures. 
Remarkably, the lowest boundary of the first MI lobes 
eventually inverts its slope as a function of the hopping rate, leading to a re-entrant scenario, 
which is maintained in 2D. We show that this effect has direct 
consequences on the spatial extension of the MI plateaux for the case of an overimposed harmonic confinement. 
Anti-intuitively, we show that the plateaux extension may become constant or even wider for 
increasing hopping.


\begin{figure}[ht] 
\begin{center}
\includegraphics[width=6.0cm]{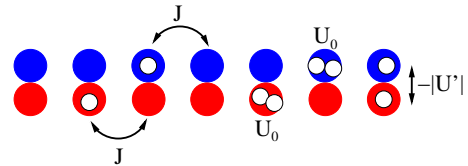}
\end{center} 
\vspace*{-0.4cm} 
\caption{Scheme of the system considered in this Letter.}  
\label{fig1}
\vspace*{-0.1cm}  
\end{figure}

In the following, we consider dipolar bosons placed at two neighboring, but disconnected, 
1D traps (wires), which can be created using micro-magnetic confinement \cite{Schmiedmayer}
or sufficiently strong 2D optical lattices. In the latter case,  
the required two-site configuration may be generated by super-lattice techniques  
or by selectively emptying 1D sites neighboring the desired pair.
Along the 1D systems we assume an additional lattice equal for both 1D traps, 
which leads to the ladder configuration shown in Fig.~\ref{fig1}. In the following we 
borrow from the literature on spin ladders 
the terms leg (each wire) and rung (pair of neighboring sites belonging to different legs).
In this Letter, we are mostly concerned about interlayer effects, and hence we  
consider a configuration for which only the (attractive) DDI between sites at the same rung
plays a significant role. This is the case, if 
the dipoles are oriented forming an angle $\varphi$ 
with the axis of the wires, such that $\cos^2\varphi=1/3$. In that case, the DDI between 
neighbors at the same leg vanishes, whereas the DDI between sites in the same
rung is attractive. There is in principle an additional non zero diagonal DDI between 
sites in neighboring rungs belonging to different legs. These terms can be made negligible by considering 
the spacing between rungs, $\gamma>1$ times larger than the separation between the two legs. 
In that case the spurious diagonal interaction is a factor 
$(1+2\sqrt{2}\gamma)/(1+\gamma^2)^{5/2}$ ($\simeq 0.03$ for $\gamma=3$) smaller 
than that between sites in the same rung. 
Of course, for other dipole and lattice configurations, 
the DDI between sites belonging to the same leg cannot be neglected, 
and interesting physics can be expected \cite{FootnoteVekua} and will be studied elsewhere.

Under the previous conditions the system is described by a Bose-Hubbard Hamiltonian (BHH) 
of the form
\begin{eqnarray}
&&\hat H=-J\sum_{\alpha=1,2}\sum_{<i,j>}\{\hat b_i^{(\alpha)\dag} \hat b_j^{(\alpha)} +H.c.\} 
-\mu \sum_{\alpha=1,2} \hat n_i^{(\alpha)}
\nonumber \\
&&+\frac{U_0}{2}\sum_{\alpha=1,2}\sum_i \hat n_i^{(\alpha)} (\hat n_i^{(\alpha)} -1)
-|U'|\sum_i \hat n_i^{(1)}\hat n_i^{(2)},
\label{BHH}
\end{eqnarray}
where $\hat b_i^{(\alpha)}$, $\hat b_i^{(\alpha)\dag}$, and $\hat n_i^{(\alpha)}$ are, respectively, 
the annihilation, creation, and number operators for the site $i$ at the leg $\alpha$. 
$J$ describes the hopping between neighboring sites $i$ and $j$ in each leg, $U_0$ 
the on-site interactions (a combination of short-range and dipolar contributions \cite{Goral}), 
and we consider the same chemical potential $\mu$ in both legs. Atoms in sites at the same rung
interact attractively by the DDI, which is characterized by a coupling $-|U'|$.

In the following we analyze the effects of the coupling $U'$ in the physics of the MI phases 
for the 1D wires. Note that the Hamiltonian (\ref{BHH}) is formally equivalent to the case of two bosonic 
species in a 1D array, with equal chemical potential $\mu$ for both, equal hopping $J$, 
equal on-site intra-species interactions $U_0$, and an interspecies interaction $-|U'|$. 
Hence, our results can be equally applied to boson-boson mixtures under these constraints. Indeed, as we 
show below, the PSF phase introduced in the context of Boson-Boson mixtures \cite{Kuklov} 
is crucial for the understanding of the physics discussed below.
\begin{figure}[ht] 
\begin{center}
\includegraphics[width=5.0cm]{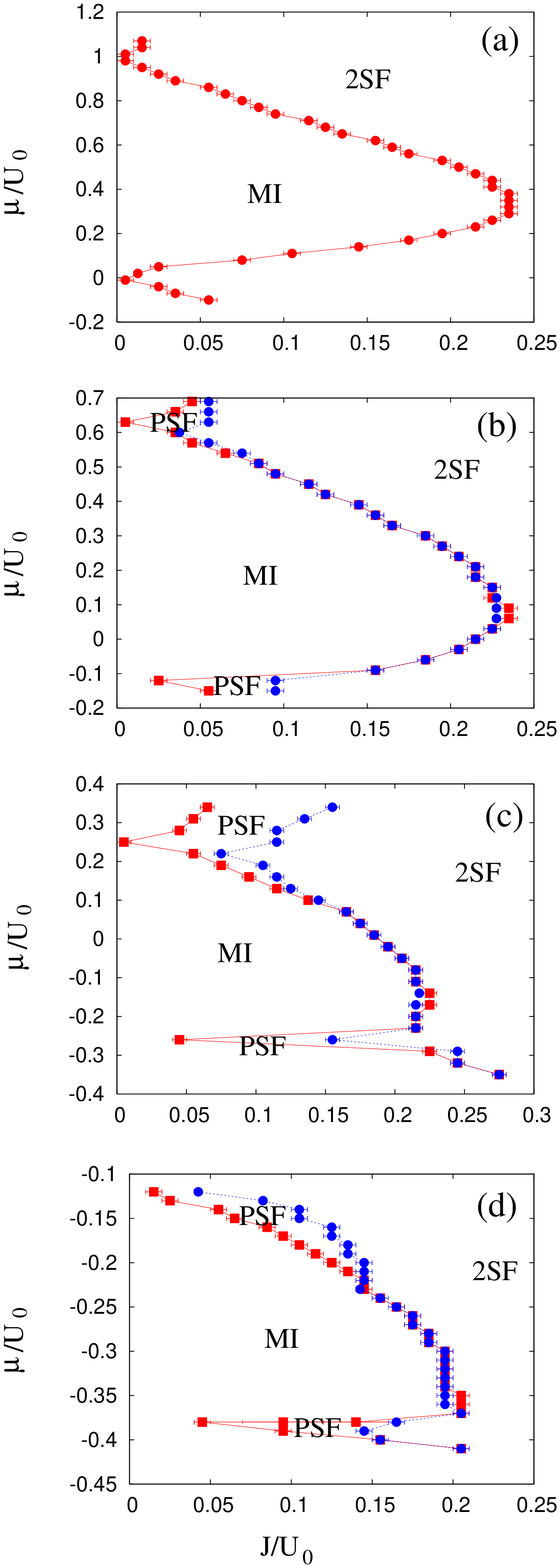}
\end{center} 
\vspace*{-0.2cm} 
\caption{MI lobes as a function of $J/U_0$ and $\mu/U_0$, 
for $|U'|/U_0=0$ (a), $0.25$ (b), $0.5$ (c) and $0.75$. 
The calculations where made for $L=12$ and $D=5$. The error bars indicate 
the change in the order parameters, 
({\color{red}$\blacksquare$})$\Psi_{PSF}$ and $({\color{blue}{\bullet}})\langle b\rangle$, 
from $<10^{-4}$ to $>10^{-2}$.}  
\label{fig2}
\vspace*{-0.1cm}  
\end{figure}

In our analysis of the ground states of $\hat H$, we have employed Matrix-Product-State (MPS) techniques, 
following closely the method of Ref.~\cite{Verstraete}. The MPS
represent an optimal Ansatz \cite{CiracOptimal} for problems 
as that of this Letter. Adapted to the two-leg problem, with $L$ sites per leg, 
the MPS Ansatz for the many-body wavefunction is:
\begin{equation}
|\Psi>=\sum_{ \{ n_i^{(\alpha)}=0\} }^{n_{max}} A_{[1]}^{[n_1^{(1)},n_1^{(2)}]}\cdots A_{[L]}^{[n_L^{(1)},n_L^{(2)}]}
|\{ n_i^{(\alpha)} \} \rangle, 
\end{equation}
where we consider a maximal number of atoms $n_{max}$ per site, and $A_{[j]}^{[n_j^{(1)},n_j^{(2)}]}$ 
is a $D\times D$ matrix, associated to the case of $n_j^{(1)}$ and $n_j^{(2)}$ atoms at the site $j$ 
of both legs. In the following we focus on the regime of low average 
occupation per site around the first MI lobe, and hence in our calculations it 
proves enough $n_{max}=2$. In addition, we have checked in our calculations that relatively 
low matrix dimensions $D=6$ describe properly the problem under consideration. The MPS Ansatz enormously 
simplify the original problem (which scales exponentially with $L$), since it has 
a complexity given by $(n_{max}+1)D^2L$. Using a similar approach as that of Ref.~\cite{Verstraete} we 
developed a numerical algorithm that allows us to recursively adapt the matrices until 
reaching the ground state. This method resembles in many ways that of finite-size 
Density-Matrix-Renormalization-Group techniques \cite{ReviewNoack}.
\begin{figure}[ht] 
\begin{center}
\includegraphics[width=5.5cm]{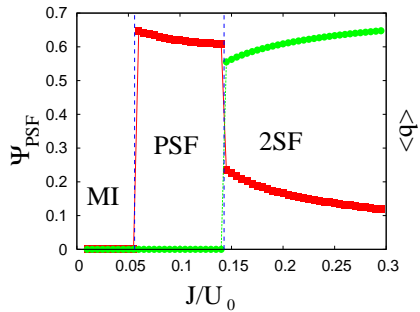}
\end{center} 
\vspace*{-0.2cm} 
\caption{({\color{red}$\blacksquare$})$\psi_{PSF}$ and $({\color{green}\bullet})\langle b \rangle$ 
as a function of $J/U_0$ for $\mu=-0.15$ and $|U'|/U_0=0.75$ with $L=24$ and $D=6$.}  
\label{fig3}
\vspace*{-0.1cm}  
\end{figure}

Fig.~\ref{fig2} shows the results of our simulations for the surroundings of the lowest MI lobe
(with $\langle n_i^{1,2}\rangle =1$) for $|U'|/U_0=0$ (a), $1/4$ (b), $1/2$ (c) and $3/4$ (d). Note 
that in order to avoid collapse in a single site, $|U'|<U_0$. 
For the case of $U'=0$, the usual Mott-lobes are recovered \cite{footnote1}. However, 
the dependence of the lobe boundaries in the $\mu$-$J$ phase space changes significantly when $|U'|$ grows. 
Note, in particular, that the lowest boundary becomes progressively flatter when $|U'|$ approaches $U_0/2$. 
Indeed our analytical calculations (see below) show that for $|U'|>U_0/2$ the slope of the lowest 
boundary of the MI lobe inverts its sign. This behavior is however difficult to observe in our 
numerical calculations due to the very narrow region between the MI-lobe 
and the region of zero occupation. In the following we discuss in more detail the physics behind  
the distortion of the MI lobes, and the implications of this distortion on the spatial extension of the 
MI lobes in axially trapped gases.

The boundaries of the MI lobes are provided by the energy gap between the MI state 
and the lowest excited state conserving the particle number. In usual (single-component) BHH \cite{Fisher89} 
this lowest excitation is provided by particle-hole excitations. The MI boundaries  
can then be calculated by a strong-coupling expansion (SCE) \cite{Monien}, estimating the energy of a state 
with an extra particle and a state with an extra hole. This is indeed the case of $U'=0$, where the lowest 
excitations are given by uncorrelated particle-hole excitations in both wires. The situation changes 
for $|U'|>0$, since for sufficiently low tunneling, there is a direct transition  between 
MI and PSF phases, i.e. superfluid phases of composites (or composite holes) \cite{Kuklov}. In that 
case the first excitation of the MI lobe is given by the correlated creation of pairs of particles (or holes) 
at opposite sites of the two wires, explaining the qualitative change in the shape of the lobe boundaries. 
In particular, a second-order SCE in $J/|U'|$ \cite{footnoteSCE}, provides the following dependence for sufficiently low tunneling 
for the lowest boundary of the MI lobe with $n_0$ particles per site:
\begin{eqnarray}
&&\frac{\mu}{U_0}=n_0-1+\frac{|U'|}{U_0}\left (\frac{1}{2}-n_0\right ) \nonumber \\
&&-4\left ( \frac{J}{U_0} \right )^2 
\left [ n_0(n_0+1)-\frac{(n_0^2-1)/2}{2-|U'|/U_0}-\frac{n_0^2U_0}{|U'|} \right ]
\label{SCE}
\end{eqnarray}
From (\ref{SCE}) it becomes clear that for any $U'>0$ the gap boundaries are quadratic (and not linear) 
in $J$ for sufficiently low $J$. Interestingly, the 
lowest boundary of the first MI region ($n_0=1$) 
inverts its slope at $J=0$ for $|U|'>U_0/2$, in agreement with our numerical results. 
One may also observe that an inversion of the slope of the lowest boundary is expected 
also for $n_0=2$ at $|U'|/U_0\simeq 0.85$, but it is not expected for $n_0>2$.

In our numerical simulations, we revealed the presence of the pairing-phases by monitoring 
$\Psi_{PSF}=|\langle \hat b^{(1)}\hat b^{(2)}\rangle - 
\langle \hat b^{(1)}\rangle \langle \hat b^{(2)}\rangle|$ \cite{footnote2}. 
A typical dependence of $\Psi_{PSF}$ and $\langle b\rangle=\langle b^{(1,2)}\rangle$ 
in our simulations is depicted in Fig.~\ref{fig3} for a fixed chemical potential. 
The MI region is characterized as that in which both  $\Psi_{PSF}=\langle b\rangle=0$. 
We denote the PSF region as that in which $\Psi_{PSF}\ne 0$ but $\langle b\rangle=0$. Finally 
the $2SF$ region is that in which $\langle b\rangle\ne 0$. Note that there is a finite coexistence region 
in which both $\Psi_{PSF}\ne 0$ and $\langle b\rangle\ne 0$. Repeating the calculations for different 
chemical potentials we obtain the results depicted in Figs.~\ref{fig2}. Note that,  
as we mentioned above, a direct MI-PSF transition can be observed in Figs.~\ref{fig2} at low $J/U_0$, which results 
in a clear distortion of the MI boundaries when compared to the $U'=0$ case. 

\begin{figure}[ht] 
\begin{center}
\includegraphics[width=8.0cm]{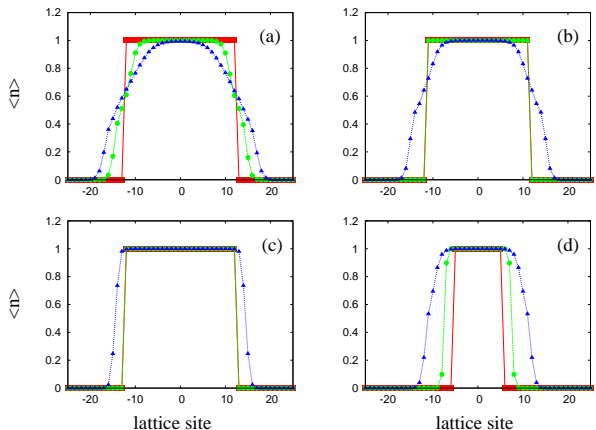}
\end{center} 
\vspace*{-0.2cm} 
\caption{Spatial distribution of $\langle n \rangle$ for $U'/U_0=0$ (a), $0.25$ (b), $0.5$ (c) and 
$0.75$ (d), for, respectively $10^4 \Omega/U_0=12.3,9.26,9.26,4.32$, with 
$\mu_0/U_0=0.2,0,-0.1,-0.36$,   
for $J/U_0=0({\color{red}\blacksquare}),0.05 ({\color{green}\bullet}),0.1 ({\color{blue}\blacktriangle})$.}  
\label{fig4}
\vspace*{-0.1cm}  
\end{figure}

The qualitative change in the shape of the MI lobe has important consequences on the 
spatial extension of the MI and SF regions in the presence of an overimposed harmonic confinement. 
In order to analyze this point, we consider a harmonic trap along the wires, 
such that a term $\Omega\sum_{i,\alpha} i^2 \hat n_i^{(\alpha)}$ is added to the BHH. 
This term induces a local chemical potential $\mu_i=\mu_0-\Omega i^2$, where $\mu_0$ is the local chemical 
potential at the trap center. Hence, $\mu_i$ scans values $\mu<\mu_0$. If for a given tunneling rate, the 
system with $\mu_0$ is inside the first MI lobe, it is hence expected the apperance of 
a MI shell at the trap center, characterized by 
a plateau in the average population per site ($\langle n \rangle=1$), surrounded 
by a second SF shell (with $\langle n \rangle < 1$) \cite{Jaksch98,Batrouni}. 
For $U'=0$, for a fixed chemical potential, it is intuitively expected that the MI plateau 
shrinks when $J/U_0$ increases, until eventually dissappear. Indeed, this is the case, since 
the lowest boundary of the MI lobe increases with $J$, hence decreasing the spatial MI region 
(Fig.~\ref{fig4}(a)). 
However, when $|U'|$ grows, the change in the slope of the lowest boundary of the first MI lobe 
leads to a significant modification of the spatial extension of the MI plateau. In particular, 
as shown in Figs.~\ref{fig4}, the basically $J$-independent lowest MI boundary 
for $|U'|=U_0/2$ leads to a $J$-independent MI plateau (Fig.~\ref{fig4}(c)). Moreover, 
for $|U'|>U_0/2$, the re-entrant character of the MI lobe leads to the anti-intuitive observation, 
that for enhanced mobility ($J$ larger) the MI plateaux become even broader (Fig.~\ref{fig4}(d)).

Although a detailed 2D analysis 
is beyond the scope of this Letter, and will be the subject of further investigations, 
we stress here that the SCE for 2D lattices at unconnected layers 
(or equivalently to two-component bosonic gases in 2D lattices) shows that the lowest boundary 
of the first MI lobe follows at low $J$ the relation~(\ref{SCE}) but substituting 
$2(J/U_0)^2$ by $z(J/U_0)^2$, where $z$ is the coordination number. Hence, the change in the sign 
of the slope occurs exactly as for 1D, and thus a 
re-entrant scenario is also expected in 2D.

In this Letter we have analyzed the physics of dipolar gases in unconnected neighboring 
1D systems. Whereas without dipolar interactions the 1D systems are independent, 
the nonlocal dipole-induced interlayer interaction leads to a direct MI to PSF transition, 
significantly distorting the MI-lobes along the wires.
In particular, the lowest boundary of the first MI lobes becomes progressively 
flatter as a function of the hopping, inverting eventually its slope, 
leading to a re-entrant configuration (that remains in 2D). 
We have shown that such an effect leads to a non-trivial behavior of the MI 
plateaux in experiments with an axial harmonic confinement \cite{BlochDomains,KetterleDomains}. 
In particular, the MI plateaux may (for low hopping) become insensitive to the hopping, or even 
anti-intuitively grow for larger tunneling. Finally, 
we would like to stress that our results also apply to two-component Bose gases, predicting 
exciting phenomenology in on-going experiments in bosonic mixtures in lattices.

\acknowledgements
Conversations with M. Lewenstein, S. We\ss el, and T. Vekua are acknowledged. 
We thank J. I. Cirac, S. Manmana, K. Rodr\'\i guez, S. We\ss el, and   
very especially J. J. Garc\'\i a -Ripoll for their support concerning the numerical simulations. 
This work was supported by the DFG (SFB-TR21, SFB407, SPP1116).


\begin{thebibliography}{99}
\bibitem{footnote1} Some features at the lobe tip are not fully recovered due to finite-size effects.
\bibitem{footnoteSCE} Note that the SCE is worse when $U'$ decreases.
\bibitem{footnote2} Strictly speaking this order parameter should vanish in the thermodynamic limit, 
since we are working in 1D. However, due 
to finite size, it acquires a finite nonzero value, being zero both in the MI phase and 
for uncorrelated pairs and holes in the two wires, and it is in this sense 
useful to characterize the appearance of the pairing phases in our calculations.
\end{thebibliography}
\end{document}